\documentclass[amsmath,amssymb,aps,twocolumn,prl]{revtex4-1}
\usepackage{graphicx}
\usepackage{dcolumn}
\usepackage{bm}

\usepackage{xcolor}
\usepackage{braket}
\definecolor{dr}{rgb}{0.8,0,0}
\definecolor{dg}{rgb}{0,0.7,0}
\definecolor{turq}{rgb}{0,0.5,0.7}

\newcommand{\spel}[1]{{\usefont{T1}{lmtt}{b}{n} #1}}
\usepackage{ulem}

\begin{document}

\title{Partial Order-Disorder Transition Driving Closure of Band Gap:\\ Example of Thermoelectric Clathrates}

\author{Maria Troppenz$^{1}$}
\email{maria.troppenz@physik.hu-berlin.de}
\author{Santiago Rigamonti$^{1}$}
\author{Jorge O. Sofo$^{2}$}
\author{Claudia Draxl$^{1}$}

\affiliation{%
$^{1}$Institut f\"ur Physik und Iris Adlershof, Humboldt-Universit{\"a}t zu Berlin, zum Gro\ss{}en Windkanal 6, 12489, Berlin, Germany \\ $^{2}$Department of Physics and Materials Research Institute, The Pennsylvania State University, University Park, Pennsylvania 16802, USA}


\date{\today}

\begin{abstract}
On the quest for efficient thermoelectrics, semiconducting behavior is a targeted property. Yet, this is often difficult to achieve due to the complex interplay between electronic structure, temperature, and  disorder. We find this to be the case for the thermoelectric clathrate Ba$_8$Al$_{16}$Si$_{30}$: Although this material exhibits a band gap in its groundstate, a temperature-driven partial order-disorder transition leads to its effective closing. This finding is enabled by a novel approach to calculate the temperature-dependent effective band structure of alloys. Our method fully accounts for the effects of short-range order and can be applied to complex alloys with many atoms in the primitive cell, without relying on effective medium approximations.
\end{abstract}

\pacs{Valid PACS appear here}
\maketitle

Disorder is a frequent phenomenon in materials used in technologically relevant applications. It affects key properties, in particular the electronic band structure that, in turn, determines electron and heat transport and more. Devising concepts for calculating the energy spectrum in the presence of disorder has a long history, including effective-medium theories such as the virtual-crystal approximation \cite{Bellaiche2000}, the coherent-potential approximation \cite{Soven1967,Soven1970}, and extensions thereof \cite{Ruban2008}. However, comprehensive consideration of systems with large numbers of atoms in their primitive cell, where local-environment effects are important, has challenged these techniques \cite{Ruban2008,Raghuraman2020}. 

In this Letter, we address this problem and demonstrate a method to compute the finite-temperature effective band structure of alloys. A combination of the cluster expansion (CE) \cite{sanchez1984} with the Wang-Landau (WL) \cite{wanglandau2001} method is employed to obtain an ensemble-averaged effective band structure from first principles. Our approach allows for uncovering the interplay between temperature-dependent alloy disorder and the related electronic properties. Atom relaxations and dopant-dopant correlations, leading to short-range order, are incorporated by construction. We demonstrate our method with the example of a complex thermoelectric alloy, namely the intermetallic clathrate Ba$_8$Al$_{16}$Si$_{30}$. For this system, we observe a temperature-driven closing of its effective band gap, concomitant with a partial order-disorder transition at $T_c=582\,$K. Since conductance by p- or n-type doping is beneficial for thermoelectricity, the appearance of metallicity in the alloy may have a degrading effect on the expected thermoelectric efficiency.

Our method enables the calculation of the electronic energy spectrum at finite temperatures from a canonical ensemble average as follows: 
\begin{itemize}
\item [(\textit{i})]
For the composition of interest, we generate a set of configurations whose energies and electronic properties are determined by {\it ab initio} calculations. The number of configurations in the set is much smaller than the actual number of available configurations.
\item [(\textit{ii})]
For each configuration in the set, we restore the symmetry of the pristine primitive cell by an averaging procedure according to Refs.~\cite{popescu2010,popescu2012,ku2010}. This symmetrized energy spectrum accounts for the different local environments present due to substitutents.
\item [(\textit{iii})]
The configurational density of states $g(E)$ is evaluated by canonical Wang-Landau sampling~\cite{wanglandau2001,wanglandau2004}. To predict the energy of the configurations visited during the sampling, we employ a CE model.
\item [(\textit{iv})]
Using $g(E)$, we assign a statistical weight to the symmetrized energy spectra from step (\textit{ii}) and perform a canonical-ensemble average. The result is the finite-temperature effective band structure. For this average, the underlying assumption is spatial ergodicity \cite{abrikosov1997}, {\it i.e.} all possible configurations in a finite sample are realized in the infinite sample. Thus, an average of many single finite samples can approximate the behavior of the infinite sample.
\end{itemize}

\begin{figure}[htb]
	\centering
	\includegraphics[width=0.35\textwidth]{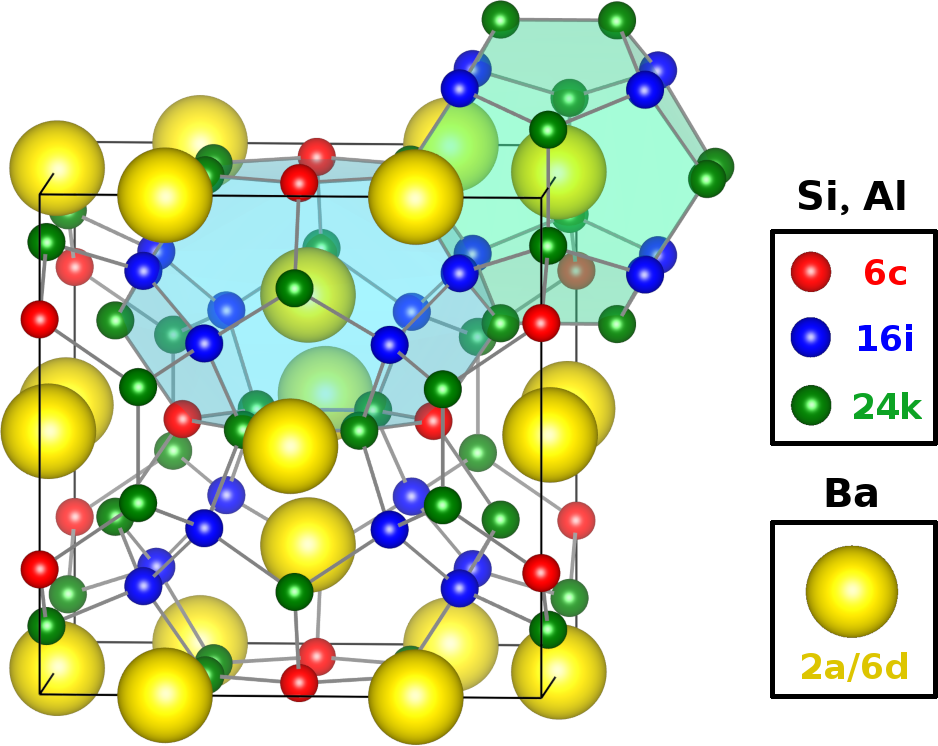}
	\caption{\label{fig:figure0} 
Unit cell of the type-I clathrate structure Ba$_8$Al$_x$Si$_{46-x}$ (space group $Pm\overline{3}n$). Host atoms are at the Wyckoff sites of the pristine lattice, $w=24k$ (green), $16i$ (blue), and $6c$ (red), guest atoms at $2a$ and $6d$ (yellow). Two guest-containing cavitites are shown in light blue and light green.}
\end{figure}

This method is applied, in the following, to calculate the finite-temperature effective band structure for the clathrate Ba$_8$Al$_{16}$Si$_{30}$. Intermetallic clathrates are inclusion compounds that encapsulate guest atoms inside cavities in their crystal lattice (see Fig.~\ref{fig:figure0}). The enormous compositional space for their synthesis offers, in principle, an excellent playground for tayloring their properties towards a high thermoelectric performance. 
The host structure of type-I clathrate compounds, consisting of 46 tetrahedrally coordinated group-IV species in the unit cell, can contain up to eight guest atoms,  often alkali or alkaline-earth metals. These act as endohedral dopants and donate their outer-shell electrons. Following the Zintl rule \cite{zintl1939}, the compound Ba$_8$Al$_{x}$Si$_{46-x}$, with Ba as guest atoms and Al atoms doping the Si host (unit cell shown in Fig.~\ref{fig:figure0}), 
is expected to be charge-balanced for $x=16$~\cite{paschen2003,christensen2010,shevelkov2011}. Indeed, previous studies have shown that the most stable structure of Ba$_8$Al$_{16}$Si$_{30}$ is a semiconductor~\cite{troppenz2017}. However, the composition is not the only factor determining the electronic properties, yet, the configuration, \textit{i.e.} the arrangement of the Al atoms in the host lattice, plays an essential role. In fact, it has been observed in several clathrate compounds, that the electronic structure is very sensitive to subtle changes in the configuration ~\cite{troppenz2017,blake1999,blake2001,akai2009,erhart2016}. In particular, configurations of Ba$_8$Al$_{16}$Si$_{30}$ with energies only a few meV/atom above the ground state are found to be metallic~\cite{troppenz2017}. This poses a challenging scenario for a reliable theoretical description of such materials at finite temperatures, since, on the one hand, the properties of a single configuration (e.g. the lowest-energy structure) cannot represent the thermal average while, on the other hand, supercell sizes needed to describe the thermodynamic limit are out of reach.

\begin{figure}[t]
	\includegraphics[width=0.45\textwidth]{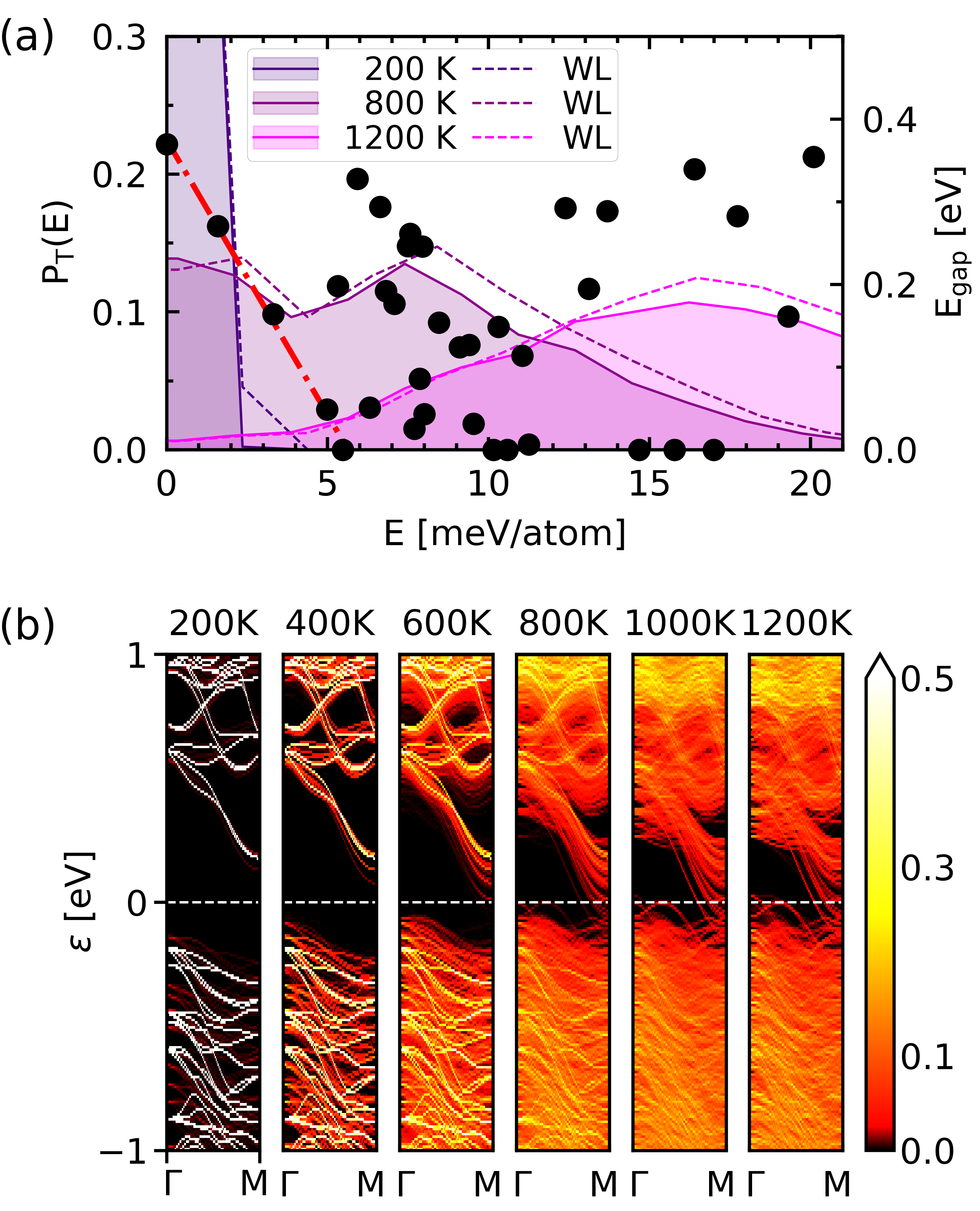}
	\caption{\label{fig:figure1}(a)
Canonical distribution $P_T(E)$ for $T=200\,$K, $800\,$K, and $1200\,$K (left axis) and Kohn-Sham band gaps $E_{\rm gap}$ (black dots, right axis) versus total energy $E$. The energy zero is the ground-state energy $E_{\rm GS}$. The results for $P_T(E)$ obtained from the MC method are shown with shaded areas, those from the WL method with dashed lines. The linear decrease of the band gap at low energies is indicated by a red dash-dotted line. (b) Spectral function along the path $\Gamma$-$M$ for temperatures between 200 and 1200 K (from left to right). The white dashed lines indicate the Fermi energy $E_{F}$. Black corresponds to $A_T (\bm{k},\epsilon)=0$, and white to $A_T (\bm{k},\epsilon)>0.5$.}
\end{figure}

To employ the above described approach, we start by calculating the electronic properties of 44 Ba$_8$Al$_{16}$Si$_{30}$ configurations \textit{ab initio} with the exchange-correlation functional PBEsol  \cite{perdew2008:PBEsol} by using the full-potential all-electron DFT package \spel{exciting} \cite{gulans2014} ({\bf step (\textit{i})}, for details see the Supplemental Material (SM)~\cite{comdetails}). In this system, the unit cell consisting of 54 atoms is already sufficiently large to account for the effects of short-range correlations in the electronic structure. The fundamental DFT band gaps, $E_{\rm gap}$, of the computed configurations with total energies below 21 meV/atom are shown in Fig.~\ref{fig:figure1}(a) (black dots, right axis) where the ground-state energy $E_{\rm GS}$ serves as a reference (see Sec.~I in SM in Ref.~\cite{comdetails}). The GS has an indirect band gap of  $0.36\,$eV~\cite{troppenz2017} along the $\Gamma$-$M$ direction. Starting from this value, the band gap decreases almost linearly for configurations of increasing energy (red dash-dotted line) until becoming zero at around $5\,$meV/atom. For larger total energies, both semiconducting and metallic structures are present.
 
In {\bf step (\textit{ii})}, we calculate the symmetrized energy spectrum in the pristine primitive cell, i.e. that of the non-substituted clathrate lattice, for each of the 44 configurations. Due to the Al substituents, the symmetry of this cell is broken, such that the eigenvalues for the wave vectors $\bm{k}$ and $\bm{k}_S=S\bm{k}$, with $S$ being a point symmetry operation of the primitive cell, are in general different. We average the energy spectrum over all point symmetry operations of the pristine lattice by defining a spectral function 
\begin{align}
	A_c(\bm{k},\epsilon)= \frac{1}{N_S} 
	\sum_{S} \sum_n \delta_{\xi} ({\epsilon - \epsilon_{c,n\bm{k}_S}}) \,. \label{eq:As}	
\end{align}
Here, $\epsilon_{c,n\bm{k}}$ is the eigenenergy of band $n$ and wave vector $\bm{k}$ for configuration $c$. $\delta_{\xi}(x)=1$ if $x\in[-\xi,\xi)$ and $0$ otherwise. $\xi$ is a small number representing the discretization of $\epsilon$ (y-axis in Fig.~\ref{fig:figure1}(b)). 
In our case, an unfolding of the bandstructure to recover the translational symmetry of the primitive cell~\cite{ku2010,popescu2010} is not required, since the lattice of the configurations has the same size as the latter.
The symmetry-averaged spectral functions are shown for the GS and for a high-energy structure in Figs.~1 and 2 of the SM \cite{comdetails}. 

Using the $A_c(\bm{k},\epsilon)$'s of the configurations with energies $E_c$, 
we can perform a canonical-ensemble average to obtain the finite temperature spectral function $A_T(\bm{k},\epsilon)$:
\begin{eqnarray}
A_T(\bm{k},\epsilon)\approx\frac{1}{Z_T}\sum_{c}A_c(\bm{k},\epsilon)e^{-E_c/k_BT}.
\label{eq:at1}
\end{eqnarray}
Here, the sum runs on the configurations $c$ of the alloy with energy $E_c$ and $Z_T$ is the canonical partition function $Z_T=\sum_ce^{-E_c/k_BT}$. $k_B$ is the Boltzmann constant. For sparse configurational samplings, the weight of different configurations in Eq.~(\ref{eq:at1}) may be misrepresented. To alleviate this problem, we employ the configurational density of states, $g(E)$, obtained by the Wang-Landau sampling~\cite{wanglandau2001,wanglandau2004} ({\bf step (\textit{iii}}), shown in Fig.~3 in the SM~\cite{comdetails}). It is calculated with the cluster-expansion package \spel{CELL}~\cite{rigamonti2020,cellURL}, using the CE model from Ref.~\cite{troppenz2017} for predicting the energy of the configurations visited in the sampling.

\begin{figure}[t]
\includegraphics[width=0.45\textwidth]{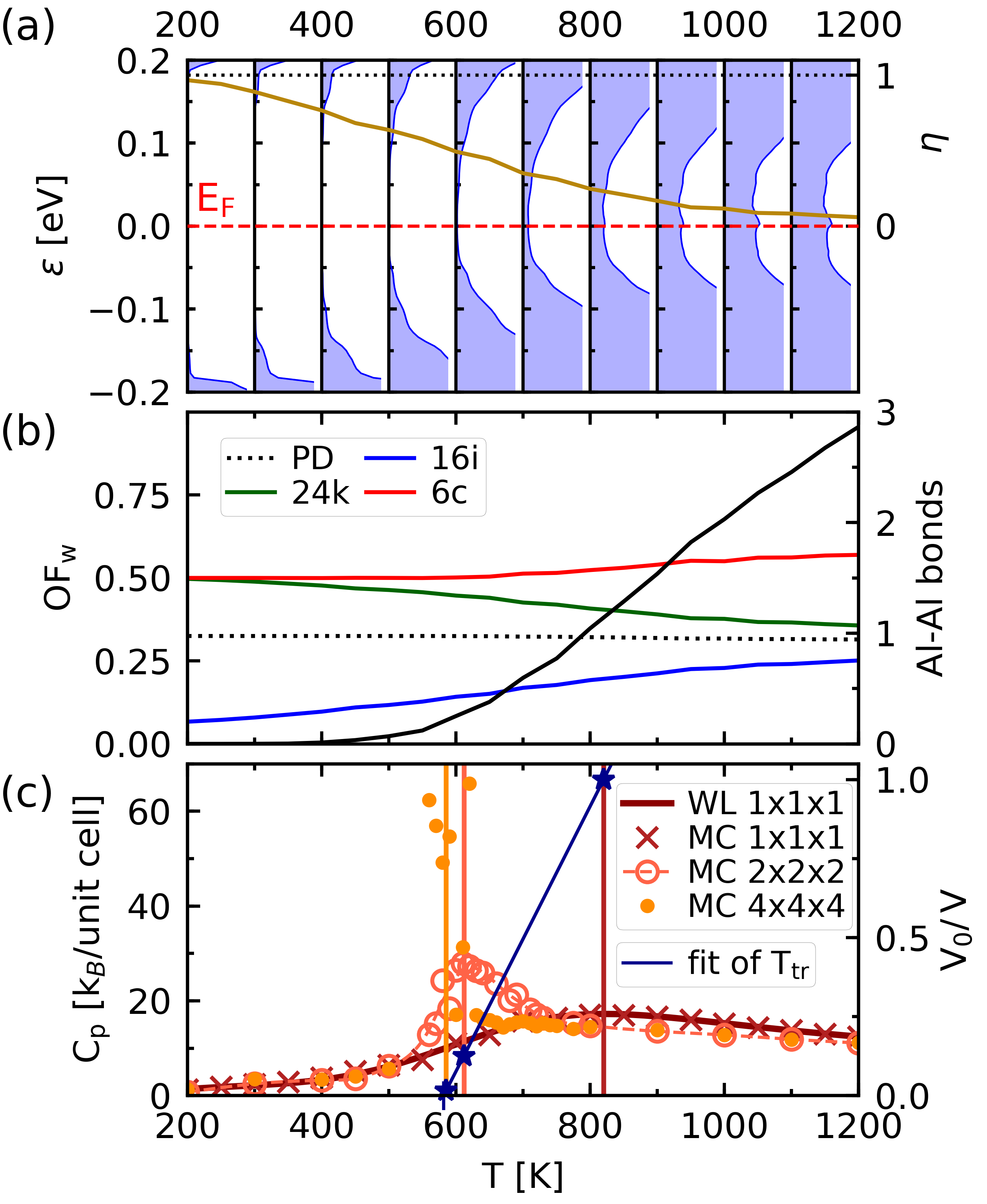}
\caption{\label{fig:figure2} Signatures of the phase transition. (a) Electronic density of states (blue shaded area, left axis; the Fermi level is indicated with a dashed red line) and order parameter $\eta$ from Eq.~(\ref{eq:eta}) (brown line, right axis); (b) occupation factors of the three Wyckoff sites $24k$, $16i$, $6c$ and ${\rm OF}_{\rm PD}$ (left axis), and number of Al-Al bonds (black line, right axis). (c) specific heat $C_p$ (left axis) for the $1\times1\times1$ cell obtained from the WL method (dark-red solid line) and from the MC method (dark-red crosses). In addition, $C_p$ is shown for the $2\times2\times2$ (dashed orange line, circles) and $4\times4\times4$ (yellow dots) supercells from MC simulations. The transition temperatures $T_{tr}$ are indicated by vertical lines of the corresponding color. The dark-blue stars indicate the inverse of the supercell volume $V^{-1}$ (right axis). The least-squares fit of Eq.~(\ref{eq:finitesize}) (dark-blue solid line) to these data points yield a slope of $a=238.4\,$K$\,V_{0}$ and the transition temperature in the macroscopic limit, $T_{tr}(\infty)=582\,K$.}
\end{figure}

After having obtained these statistical weights, $g(E)$, as a function of the energy $E$, the sum over the configurations $c$ in Eq.~(\ref{eq:at1}) can be recast into a sum over energy intervals as (see Sec.~II in SM~\cite{comdetails})
\begin{align}\label{eq:AT}
	A_T(\bm{k},\epsilon)= \, \frac{1}{Z_T}\sum_{i=0}^M \,
	\langle A_c(\bm{k},\epsilon) \rangle_i
	\Delta_i g(E_i)e^{-E_i/k_B T}.
\end{align} 
This canonical-ensemble average yields the energy spectrum at finite temperature ({\bf step (\textit{iv})}). Here, $\langle A_c(\bm{k},\epsilon) \rangle_i= \sum_{c\in\Delta_{E_i}} A_c(\bm{k},\epsilon)/n_i$ is the configuration-averaged spectral weight in the interval $\Delta_{E_i}=[E_i,E_i+\Delta_i)$, with $\Delta_i$ being small interval widths. $n_i$ is the number of computed configurations in $\Delta_{E_i}$, while the total number of configurations in the same interval is $\Delta_i g(E_i)$, and $Z_T=\sum_i^{M} \Delta_i g(E_i) e^{-E_i /{k_B T}}$. The finite-temperature effective band structure resulting from the $A_T(\bm{k},\epsilon)$'s is shown in Fig.~\ref{fig:figure1}(b). At $T\leq400\,$K, there is a small indirect effective band gap between the valence band maximum close to the $\Gamma$ point and the conduction band minimum at the $M$ point. With increasing temperature, the effective band gap starts to decrease, and above $600\,$K, the spectral function at the Fermi energy becomes non zero. This indicates a possible metallic state, given that no electron localization occurs (and we note that such effects are beyond the current theoretical description). The narrowing of the effective band gap with increasing temperature is also evident from the temperature-dependent density of states, as defined in Ref.~\cite{comdetails}, which is shown for $200\,$K~$\leq T \leq 1200\,$K in Fig.~\ref{fig:figure2}(a). Here, the effective band gap closes at around $700\,$K.

It is interesting to explore the structural and thermodynamic properties along this transition of band-gap closure. To this extent, we use the CE model of Ref.~\onlinecite{troppenz2017} to perform a configurational thermodynamics analysis of the system by means of finite-temperature canonical Metropolis Monte Carlo (MC) simulations~\cite{metropolis1953,hastings1970}. 
Figure.~\ref{fig:figure1}(a), left axis, shows the canonical probability distribution from MC (solid lines). For $T=200\,$K, $P_T(E)$ has a single peak below $3\,$meV/atom. In this energy range, only semiconducting configurations are present. With increasing temperature, $P_T(E)$ becomes more pronounced in the region with metallic configurations (above $5\,$meV/atom). At $800\,$K, the probability distribution shows two maxima, one around $2\,$meV/atom and one around $8\,$meV/atom, signaling the coexistence of two phases. For $T=1200$\,K, the distribution becomes broader with a single maximum at high energies. Similar results are obtained from the WL method using the expression $P_T(E)=  g(E)e^{-E/k_B T}/Z_T$ (dashed lines).

%

To assess the dopant configurations, the Al occupancy factors, ${\rm OF}_w (T)$, are shown in Fig.~\ref{fig:figure2}(b). ${\rm OF}_w (T)$ is defined as the fractional number of Al atoms at the Wyckoff site $w$, with $w=24k,16i$, or $6c$ (see Fig.~\ref{fig:figure0}). At $T\leq 200\,$K, the ${\rm OF}$s are
almost identical to those of the ground-state configuration, that has twelve, one, and three Al atoms at the $24k$, $16i$ and $6c$ site,
respectively (\textit{i.e.} ${\rm OF}_{24k}=0.5$, ${\rm OF}_{16i}=0.0625$, and ${\rm OF}_{6c}=0.5$), and lacks Al-Al bonds~\cite{troppenz2017}. ${\rm OF}_{6c}$ remains almost constant over the full temperature range. At $T=1200\,$K  ${\rm OF}_{6c}=0.57$, i.e. $\sim 0.4\,$Al atoms more than the GS. In contrast, ${\rm OF}_{24k}$ decreases and ${\rm OF}_{16i}$ increases with temperature. At $T \approx 1200\,$K, the number of Al-Al bonds has increased considerably (black line), where approximately $3.5$ Al atoms have been transferred from the $24k$ to the $16i$ sublattice, approaching the value of partial disorder, calculated as ${\rm OF}_{\rm PD} (T)=[16-{\rm OF}_{6c}(T)\cdot6]/40$ (black-dotted line in Fig.~\ref{fig:figure2}(b)). This value corresponds to a partially disordered structure, in which the sublattice formed by the $24k$ and $16i$ positions hosts a fully random Al-Si alloy. 
Since the actual Al content in the $24k$-$16i$ sublattice depends on the occupation of the $6c$ site, ${\rm OF}_{\rm PD}(T)$ is slightly temperature dependent, changing from $0.325$ at $200\,$K to $0.315$ at $1200\,$K.

The phase transition from a well-ordered state at low temperatures to a partially disordered (PD) state at high temperatures can be characterized by an order parameter~\cite{bragg1934,bethe1935} that we define such to be able to distinguish between these two phases
\begin{align}\label{eq:eta}
	\eta(T) = \frac{1}{2} \sum_{w=24k,16i} \left[ \frac{{\rm OF}_w(T)-{\rm
			OF}_{\rm
			PD}(T)}{{\rm OF}_w(0\,{\rm K})-{\rm OF}_{\rm PD}(0\,{\rm K})}
	\right]^2\,.
\end{align} 
Per definition, it is exactly one for the ordered phase at zero Kelvin, and becomes zero for the perfectly PD phase. Accordingly, it decreases from $\sim$1 at 200K to almost zero at 1200K, as shown in Fig.~\ref{fig:figure2}(a) (brown line).

To determine the transition temperature $T_{tr}$, we investigate its signatures on the canonical probability distribution $P_T(E)$ and the isobaric specific heat $C_p(T)=\left( \langle E^2 \rangle - \langle E \rangle^2 \right)/k_B T^2$. For first-order phase transitions, $P_T(E)$ at $T_{tr}$ is expected to display two peaks with equal height~\cite{wanglandau2001,landau2014}, and $C_p$ is expected to exhibit a maximum at the transition temperature~\cite{landau2014,challa1986,binder1987,binder1997}. As evident from Fig.~\ref{fig:figure1}(a), a double peak structure in $P_T(E)$ with nearly equal peak heights is, indeed, observed for a temperature of about $800\,$K. The results for $C_p(T)$ are shown in Fig.~\ref{fig:figure2}(c) for simulations using different supercell sizes. For a 1$\times$1$\times$1 cell, $C_p$ has a maximum at $820\,$K (indicated by the dark-red vertical line), and both WL and MC simulations yield indistinguishable results.

Due to the finite size of the simulation cell, the computed value of $T_{tr}$ deviates systematically from the macroscopic limit~\cite{challa1986,landau2014}. The transition temperature is expected to change linearly with the inverse of the simulation-cell volume $V^{-1}$~\cite{landau2014,challa1986} as 
\begin{align} \label{eq:finitesize}
	T_{tr} ( V )= T_{tr}(\infty) + a \, V^{-1} ~.
\end{align}
Here, $T_{tr}(\infty)$ is the transition temperature in the macroscopic limit, and $a$ is a constant. To determine $T_{tr}(\infty)$, we calculate $T_{tr}$ for a $2\times2\times2$ and a $4\times4\times4$ supercell, i.e., $V=8\,V_0$ and $64\,V_0$, respectively ($V_0$ being the unit-cell volume), and fit Eq.~(\ref{eq:finitesize}) to these data points. For the $2\times2\times2$ supercell, we obtain $T_{tr}=621\,$K from the maximum of $C_p$ (see Fig. \ref{fig:figure2}(c), circles). For increasing supercell sizes, very large times are required to sample the distribution effectively close to the transition temperature~\cite{newman1999,landau2014,challa1986}. This leads to an increased uncertainty of the computed $C_p$ values, as can be seen for the $4\times4\times4$ supercell around $T=600\,$K (Fig. \ref{fig:figure2}(c), dots), hampering an accurate determination of the peak position. Nonetheless, from a direct inspection of the MC trajectories for temperatures between 560K and 610K, we see that the bimodal character of $g(E)$ leads to a sudden jump between low-energy and high-energy configurations at $T_{tr}\approx585\,$K ~(see Fig.~4 of SM \cite{comdetails}). We take this value as a reasonable estimate of $T_{tr}$ for the $4\times4\times4$ supercell. The respective values of $T_{tr}$ are indicated by vertical lines in Fig.~\ref{fig:figure2}(c). By performing a least-squares fit of Eq.~(\ref{eq:finitesize}) to the $T_{tr}$'s for the three supercell sizes (Fig.~\ref{fig:figure2}(c), dark-blue solid line), we obtain $T_{tr}(\infty)=582\,$K. This temperature differs significantly from $T_{tr}$ of the unit cell, emphasizing the need of a finite-size scaling. We observe furthermore that  with increasing cell size the peak of $C_p$ increases in height and decreases in width, as typically observed for first-order phase transitions~\cite{binder1987,landau2014}.

A final cross-check for a first-order transition is the temperature-dependent behavior of the entropy, where an inflection point at $T_{tr}$ in the microcanonical ensemble is a necessary condition~\cite{schnabel2011}. We, indeed, verify this behavior as obvious in Fig.~5 of the SM~\cite{comdetails}.


To summarize, we have developed a method to obtain the temperature dependent effective band-structure of alloys. Our method can be applied to complex systems with many atoms in the primitive cell. Local atomic environments are fully accounted. 
Thanks to a symmetrization procedure, the resulting spectral function could be compared to angle-resolved photoemission spectra (ARPES)~\cite{ku2010}. We have challenged our method by applying it to the thermoelectric clathrate alloy Ba$_8$Al$_{16}$Si$_{30}$. The configurational changes of Al atoms in the Si host structure as a function of temperature reveal a partial order-disorder phase transition. The critical temperature of this phase transition in the macroscopic limit is determined as $582\,$K. This transition goes hand in hand with a closing of the effective band gap, which is expected to dramatically impact the thermoelectric efficiency. As a consequence, it is anticipated that clathrate phases, annealed at different temperatures, exhibit large differences in their thermoelectric performance. Our findings point to the crucial role of disorder in complex thermoelectric materials. Overall, we have demonstrated that a multi-scale approach is needed to obtain a reliable description of the macroscopic properties for such complex materials. In particular, it is essential to capture the diverse temperature-dependent configurational effects present in those alloys rather than restricting calculations to ground-state properties and/or a few selected structures. We have further shown that a finite-size scaling is required to reach the macroscopic description at the critical point of the material's phase transition.

M.T. acknowledges funding from the Elsa-Neumann Stiftung Berlin. We thank Oleg Rubel for drawing our attention to Refs.~\cite{popescu2010,popescu2012}. Input and output files can be downloaded from the NOMAD Repository by the following link \url{http://dx.doi.org/10.17172/NOMAD/2019.10.29-1 }.


\begin{thebibliography}{37}%
	\makeatletter
	\providecommand \@ifxundefined [1]{%
		\@ifx{#1\undefined}
	}%
	\providecommand \@ifnum [1]{%
		\ifnum #1\expandafter \@firstoftwo
		\else \expandafter \@secondoftwo
		\fi
	}%
	\providecommand \@ifx [1]{%
		\ifx #1\expandafter \@firstoftwo
		\else \expandafter \@secondoftwo
		\fi
	}%
	\providecommand \natexlab [1]{#1}%
	\providecommand \enquote  [1]{``#1''}%
	\providecommand \bibnamefont  [1]{#1}%
	\providecommand \bibfnamefont [1]{#1}%
	\providecommand \citenamefont [1]{#1}%
	\providecommand \href@noop [0]{\@secondoftwo}%
	\providecommand \href [0]{\begingroup \@sanitize@url \@href}%
	\providecommand \@href[1]{\@@startlink{#1}\@@href}%
	\providecommand \@@href[1]{\endgroup#1\@@endlink}%
	\providecommand \@sanitize@url [0]{\catcode `\\12\catcode `\$12\catcode
		`\&12\catcode `\#12\catcode `\^12\catcode `\_12\catcode `\%12\relax}%
	\providecommand \@@startlink[1]{}%
	\providecommand \@@endlink[0]{}%
	\providecommand \url  [0]{\begingroup\@sanitize@url \@url }%
	\providecommand \@url [1]{\endgroup\@href {#1}{\urlprefix }}%
	\providecommand \urlprefix  [0]{URL }%
	\providecommand \Eprint [0]{\href }%
	\providecommand \doibase [0]{http://dx.doi.org/}%
	\providecommand \selectlanguage [0]{\@gobble}%
	\providecommand \bibinfo  [0]{\@secondoftwo}%
	\providecommand \bibfield  [0]{\@secondoftwo}%
	\providecommand \translation [1]{[#1]}%
	\providecommand \BibitemOpen [0]{}%
	\providecommand \bibitemStop [0]{}%
	\providecommand \bibitemNoStop [0]{.\EOS\space}%
	\providecommand \EOS [0]{\spacefactor3000\relax}%
	\providecommand \BibitemShut  [1]{\csname bibitem#1\endcsname}%
	\let\auto@bib@innerbib\@empty
	\bibitem [{\citenamefont {Bellaiche}\ and\ \citenamefont
		{Vanderbilt}(2000)}]{Bellaiche2000}%
	\BibitemOpen
	\bibfield  {author} {\bibinfo {author} {\bibfnamefont {L.}~\bibnamefont
			{Bellaiche}}\ and\ \bibinfo {author} {\bibfnamefont {D.}~\bibnamefont
			{Vanderbilt}},\ }\href@noop {} {\bibfield  {journal} {\bibinfo  {journal}
			{Phys. Rev. B}\ }\textbf {\bibinfo {volume} {61}},\ \bibinfo {pages} {7877}
		(\bibinfo {year} {2000})}\BibitemShut {NoStop}%
	\bibitem [{\citenamefont {Soven}(1967)}]{Soven1967}%
	\BibitemOpen
	\bibfield  {author} {\bibinfo {author} {\bibfnamefont {P.}~\bibnamefont
			{Soven}},\ }\href@noop {} {\bibfield  {journal} {\bibinfo  {journal} {Phys.
				Rev.}\ }\textbf {\bibinfo {volume} {156}},\ \bibinfo {pages} {809} (\bibinfo
		{year} {1967})}\BibitemShut {NoStop}%
	\bibitem [{\citenamefont {Soven}(1970)}]{Soven1970}%
	\BibitemOpen
	\bibfield  {author} {\bibinfo {author} {\bibfnamefont {P.}~\bibnamefont
			{Soven}},\ }\href@noop {} {\bibfield  {journal} {\bibinfo  {journal} {Phys.
				Rev. B}\ }\textbf {\bibinfo {volume} {2}},\ \bibinfo {pages} {4715} (\bibinfo
		{year} {1970})}\BibitemShut {NoStop}%
	\bibitem [{\citenamefont {Ruban}\ and\ \citenamefont
		{Abrikosov}(2008)}]{Ruban2008}%
	\BibitemOpen
	\bibfield  {author} {\bibinfo {author} {\bibfnamefont {A.~V.}\ \bibnamefont
			{Ruban}}\ and\ \bibinfo {author} {\bibfnamefont {I.~A.}\ \bibnamefont
			{Abrikosov}},\ }\href@noop {} {\bibfield  {journal} {\bibinfo  {journal}
			{Reports on Progress in Physics}\ }\textbf {\bibinfo {volume} {71}},\
		\bibinfo {pages} {046501} (\bibinfo {year} {2008})}\BibitemShut {NoStop}%
	\bibitem [{\citenamefont {Raghuraman}\ \emph {et~al.}(2020)\citenamefont
		{Raghuraman}, \citenamefont {Wang},\ and\ \citenamefont
		{Widom}}]{Raghuraman2020}%
	\BibitemOpen
	\bibfield  {author} {\bibinfo {author} {\bibfnamefont {V.}~\bibnamefont
			{Raghuraman}}, \bibinfo {author} {\bibfnamefont {Y.}~\bibnamefont {Wang}}, \
		and\ \bibinfo {author} {\bibfnamefont {M.}~\bibnamefont {Widom}},\
	}\href@noop {} {\bibfield  {journal} {\bibinfo  {journal} {Phys. Rev. B}\
		}\textbf {\bibinfo {volume} {102}},\ \bibinfo {pages} {054207} (\bibinfo
		{year} {2020})}\BibitemShut {NoStop}%
	\bibitem [{\citenamefont {Sanchez}\ \emph {et~al.}(1984)\citenamefont
		{Sanchez}, \citenamefont {Ducastelle},\ and\ \citenamefont
		{Gratias}}]{sanchez1984}%
	\BibitemOpen
	\bibfield  {author} {\bibinfo {author} {\bibfnamefont {J.~M.}\ \bibnamefont
			{Sanchez}}, \bibinfo {author} {\bibfnamefont {F.}~\bibnamefont {Ducastelle}},
		\ and\ \bibinfo {author} {\bibfnamefont {D.}~\bibnamefont {Gratias}},\
	}\href@noop {} {\bibfield  {journal} {\bibinfo  {journal} {Physica A}\
		}\textbf {\bibinfo {volume} {128}},\ \bibinfo {pages} {334} (\bibinfo {year}
		{1984})}\BibitemShut {NoStop}%
	\bibitem [{\citenamefont {Wang}\ and\ \citenamefont
		{Landau}(2001)}]{wanglandau2001}%
	\BibitemOpen
	\bibfield  {author} {\bibinfo {author} {\bibfnamefont {F.}~\bibnamefont
			{Wang}}\ and\ \bibinfo {author} {\bibfnamefont {D.~P.}\ \bibnamefont
			{Landau}},\ }\href@noop {} {\bibfield  {journal} {\bibinfo  {journal}
			{Physical Review Letters}\ }\textbf {\bibinfo {volume} {86}},\ \bibinfo
		{pages} {2050} (\bibinfo {year} {2001})}\BibitemShut {NoStop}%
	\bibitem [{\citenamefont {Popescu}\ and\ \citenamefont
		{Zunger}(2010)}]{popescu2010}%
	\BibitemOpen
	\bibfield  {author} {\bibinfo {author} {\bibfnamefont {V.}~\bibnamefont
			{Popescu}}\ and\ \bibinfo {author} {\bibfnamefont {A.}~\bibnamefont
			{Zunger}},\ }\href {\doibase 10.1103/PhysRevLett.104.236403} {\bibfield
		{journal} {\bibinfo  {journal} {Phys. Rev. Lett.}\ }\textbf {\bibinfo
			{volume} {104}},\ \bibinfo {pages} {236403} (\bibinfo {year}
		{2010})}\BibitemShut {NoStop}%
	\bibitem [{\citenamefont {Popescu}\ and\ \citenamefont
		{Zunger}(2012)}]{popescu2012}%
	\BibitemOpen
	\bibfield  {author} {\bibinfo {author} {\bibfnamefont {V.}~\bibnamefont
			{Popescu}}\ and\ \bibinfo {author} {\bibfnamefont {A.}~\bibnamefont
			{Zunger}},\ }\href {\doibase 10.1103/PhysRevB.85.085201} {\bibfield
		{journal} {\bibinfo  {journal} {Phys. Rev. B}\ }\textbf {\bibinfo {volume}
			{85}},\ \bibinfo {pages} {085201} (\bibinfo {year} {2012})}\BibitemShut
	{NoStop}%
	\bibitem [{\citenamefont {Ku}\ \emph {et~al.}(2010)\citenamefont {Ku},
		\citenamefont {Berlijn},\ and\ \citenamefont {Lee}}]{ku2010}%
	\BibitemOpen
	\bibfield  {author} {\bibinfo {author} {\bibfnamefont {W.}~\bibnamefont
			{Ku}}, \bibinfo {author} {\bibfnamefont {T.}~\bibnamefont {Berlijn}}, \ and\
		\bibinfo {author} {\bibfnamefont {C.-C.}\ \bibnamefont {Lee}},\ }\href@noop
	{} {\bibfield  {journal} {\bibinfo  {journal} {Phys. Rev. Lett.}\ }\textbf
		{\bibinfo {volume} {104}},\ \bibinfo {pages} {216401} (\bibinfo {year}
		{2010})}\BibitemShut {NoStop}%
	\bibitem [{\citenamefont {Landau}\ \emph {et~al.}(2004)\citenamefont {Landau},
		\citenamefont {Tsai},\ and\ \citenamefont {Exler}}]{wanglandau2004}%
	\BibitemOpen
	\bibfield  {author} {\bibinfo {author} {\bibfnamefont {D.~P.}\ \bibnamefont
			{Landau}}, \bibinfo {author} {\bibfnamefont {S.-H.}\ \bibnamefont {Tsai}}, \
		and\ \bibinfo {author} {\bibfnamefont {M.}~\bibnamefont {Exler}},\ }\href
	{\doibase 10.1119/1.1707017} {\bibfield  {journal} {\bibinfo  {journal}
			{American Journal of Physics}\ }\textbf {\bibinfo {volume} {72}},\ \bibinfo
		{pages} {1294} (\bibinfo {year} {2004})}\BibitemShut {NoStop}%
	\bibitem [{\citenamefont {Abrikosov}\ \emph {et~al.}(1997)\citenamefont
		{Abrikosov}, \citenamefont {Simak}, \citenamefont {Johansson}, \citenamefont
		{Ruban},\ and\ \citenamefont {Skriver}}]{abrikosov1997}%
	\BibitemOpen
	\bibfield  {author} {\bibinfo {author} {\bibfnamefont {I.~A.}\ \bibnamefont
			{Abrikosov}}, \bibinfo {author} {\bibfnamefont {S.~I.}\ \bibnamefont
			{Simak}}, \bibinfo {author} {\bibfnamefont {B.}~\bibnamefont {Johansson}},
		\bibinfo {author} {\bibfnamefont {A.~V.}\ \bibnamefont {Ruban}}, \ and\
		\bibinfo {author} {\bibfnamefont {H.~L.}\ \bibnamefont {Skriver}},\
	}\href@noop {} {\bibfield  {journal} {\bibinfo  {journal} {Phys. Rev. B}\
		}\textbf {\bibinfo {volume} {56}},\ \bibinfo {pages} {9319} (\bibinfo {year}
		{1997})}\BibitemShut {NoStop}%
	\bibitem [{\citenamefont {Zintl}(1939)}]{zintl1939}%
	\BibitemOpen
	\bibfield  {author} {\bibinfo {author} {\bibfnamefont {E.}~\bibnamefont
			{Zintl}},\ }\href {\doibase 10.1002/ange.19390520102} {\bibfield  {journal}
		{\bibinfo  {journal} {Angewandte Chemie}\ }\textbf {\bibinfo {volume} {52}},\
		\bibinfo {pages} {1} (\bibinfo {year} {1939})}\BibitemShut {NoStop}%
	\bibitem [{\citenamefont {Paschen}\ \emph {et~al.}(2003)\citenamefont
		{Paschen}, \citenamefont {Pacheco}, \citenamefont {Bentien}, \citenamefont
		{Sanchez}, \citenamefont {Carrillo-Cabrera}, \citenamefont {Baenitz},
		\citenamefont {Iversen}, \citenamefont {Grin},\ and\ \citenamefont
		{Steglich}}]{paschen2003}%
	\BibitemOpen
	\bibfield  {author} {\bibinfo {author} {\bibfnamefont {S.}~\bibnamefont
			{Paschen}}, \bibinfo {author} {\bibfnamefont {V.}~\bibnamefont {Pacheco}},
		\bibinfo {author} {\bibfnamefont {A.}~\bibnamefont {Bentien}}, \bibinfo
		{author} {\bibfnamefont {A.}~\bibnamefont {Sanchez}}, \bibinfo {author}
		{\bibfnamefont {W.}~\bibnamefont {Carrillo-Cabrera}}, \bibinfo {author}
		{\bibfnamefont {M.}~\bibnamefont {Baenitz}}, \bibinfo {author} {\bibfnamefont
			{B.}~\bibnamefont {Iversen}}, \bibinfo {author} {\bibfnamefont
			{Y.}~\bibnamefont {Grin}}, \ and\ \bibinfo {author} {\bibfnamefont
			{F.}~\bibnamefont {Steglich}},\ }\href@noop {} {\bibfield  {journal}
		{\bibinfo  {journal} {Physica B: Condensed Matter}\ }\textbf {\bibinfo
			{volume} {328}},\ \bibinfo {pages} {39 } (\bibinfo {year} {2003})},\ \bibinfo
	{note} {proceedings of the Second Hiroshima Workshop on Transport and Thermal
		Properties of Advanced Materials}\BibitemShut {NoStop}%
	\bibitem [{\citenamefont {Christensen}\ \emph {et~al.}(2010)\citenamefont
		{Christensen}, \citenamefont {Johnsen},\ and\ \citenamefont
		{Iversen}}]{christensen2010}%
	\BibitemOpen
	\bibfield  {author} {\bibinfo {author} {\bibfnamefont {M.}~\bibnamefont
			{Christensen}}, \bibinfo {author} {\bibfnamefont {S.}~\bibnamefont
			{Johnsen}}, \ and\ \bibinfo {author} {\bibfnamefont {B.~B.}\ \bibnamefont
			{Iversen}},\ }\href {\doibase 10.1039/B916400F} {\bibfield  {journal}
		{\bibinfo  {journal} {Dalton Trans.}\ }\textbf {\bibinfo {volume} {39}},\
		\bibinfo {pages} {978} (\bibinfo {year} {2010})}\BibitemShut {NoStop}%
	\bibitem [{\citenamefont {Shevelkov}\ and\ \citenamefont
		{Kovnir}(2011)}]{shevelkov2011}%
	\BibitemOpen
	\bibfield  {author} {\bibinfo {author} {\bibfnamefont {A.~V.}\ \bibnamefont
			{Shevelkov}}\ and\ \bibinfo {author} {\bibfnamefont {K.}~\bibnamefont
			{Kovnir}},\ }\enquote {\bibinfo {title} {Zintl phases: {P}rinciples and
			{R}ecent {D}evelopments},}\ \ (\bibinfo  {publisher} {Springer Berlin
		Heidelberg},\ \bibinfo {address} {Berlin, Heidelberg},\ \bibinfo {year}
	{2011})\ pp.\ \bibinfo {pages} {97--142}\BibitemShut {NoStop}%
	\bibitem [{\citenamefont {Troppenz}\ \emph {et~al.}(2017)\citenamefont
		{Troppenz}, \citenamefont {Rigamonti},\ and\ \citenamefont
		{Draxl}}]{troppenz2017}%
	\BibitemOpen
	\bibfield  {author} {\bibinfo {author} {\bibfnamefont {M.}~\bibnamefont
			{Troppenz}}, \bibinfo {author} {\bibfnamefont {S.}~\bibnamefont {Rigamonti}},
		\ and\ \bibinfo {author} {\bibfnamefont {C.}~\bibnamefont {Draxl}},\
	}\href@noop {} {\bibfield  {journal} {\bibinfo  {journal} {Chem. Mater.}\
		}\textbf {\bibinfo {volume} {29}},\ \bibinfo {pages} {2414} (\bibinfo {year}
		{2017})}\BibitemShut {NoStop}%
	\bibitem [{\citenamefont {Blake}\ \emph {et~al.}(1999)\citenamefont {Blake},
		\citenamefont {M{\o}llnitz}, \citenamefont {Kresse},\ and\ \citenamefont
		{Metiu}}]{blake1999}%
	\BibitemOpen
	\bibfield  {author} {\bibinfo {author} {\bibfnamefont {N.~P.}\ \bibnamefont
			{Blake}}, \bibinfo {author} {\bibfnamefont {L.}~\bibnamefont {M{\o}llnitz}},
		\bibinfo {author} {\bibfnamefont {G.}~\bibnamefont {Kresse}}, \ and\ \bibinfo
		{author} {\bibfnamefont {H.}~\bibnamefont {Metiu}},\ }\href {\doibase
		10.1063/1.479615} {\bibfield  {journal} {\bibinfo  {journal} {The Journal of
				Chemical Physics}\ }\textbf {\bibinfo {volume} {111}},\ \bibinfo {pages}
		{3133} (\bibinfo {year} {1999})}\BibitemShut {NoStop}%
	\bibitem [{\citenamefont {Blake}\ \emph {et~al.}(2001)\citenamefont {Blake},
		\citenamefont {Bryan}, \citenamefont {Latturner}, \citenamefont
		{M{\o}llnitz}, \citenamefont {Stucky},\ and\ \citenamefont
		{Metiu}}]{blake2001}%
	\BibitemOpen
	\bibfield  {author} {\bibinfo {author} {\bibfnamefont {N.~P.}\ \bibnamefont
			{Blake}}, \bibinfo {author} {\bibfnamefont {D.}~\bibnamefont {Bryan}},
		\bibinfo {author} {\bibfnamefont {S.}~\bibnamefont {Latturner}}, \bibinfo
		{author} {\bibfnamefont {L.}~\bibnamefont {M{\o}llnitz}}, \bibinfo {author}
		{\bibfnamefont {G.~D.}\ \bibnamefont {Stucky}}, \ and\ \bibinfo {author}
		{\bibfnamefont {H.}~\bibnamefont {Metiu}},\ }\href@noop {} {\bibfield
		{journal} {\bibinfo  {journal} {Journal of Chemical Physics}\ }\textbf
		{\bibinfo {volume} {114}},\ \bibinfo {pages} {10063} (\bibinfo {year}
		{2001})}\BibitemShut {NoStop}%
	\bibitem [{\citenamefont {Akai}\ \emph {et~al.}(2009)\citenamefont {Akai},
		\citenamefont {Uemura}, \citenamefont {Kishimoto}, \citenamefont {Tanaka},
		\citenamefont {Kurisu}, \citenamefont {Yamamoto}, \citenamefont {Koyanagi},
		\citenamefont {Koga}, \citenamefont {Anno},\ and\ \citenamefont
		{Matsuura}}]{akai2009}%
	\BibitemOpen
	\bibfield  {author} {\bibinfo {author} {\bibfnamefont {K.}~\bibnamefont
			{Akai}}, \bibinfo {author} {\bibfnamefont {T.}~\bibnamefont {Uemura}},
		\bibinfo {author} {\bibfnamefont {K.}~\bibnamefont {Kishimoto}}, \bibinfo
		{author} {\bibfnamefont {T.}~\bibnamefont {Tanaka}}, \bibinfo {author}
		{\bibfnamefont {H.}~\bibnamefont {Kurisu}}, \bibinfo {author} {\bibfnamefont
			{S.}~\bibnamefont {Yamamoto}}, \bibinfo {author} {\bibfnamefont
			{T.}~\bibnamefont {Koyanagi}}, \bibinfo {author} {\bibfnamefont
			{K.}~\bibnamefont {Koga}}, \bibinfo {author} {\bibfnamefont {H.}~\bibnamefont
			{Anno}}, \ and\ \bibinfo {author} {\bibfnamefont {M.}~\bibnamefont
			{Matsuura}},\ }\href {\doibase 10.1007/s11664-009-0727-1} {\bibfield
		{journal} {\bibinfo  {journal} {Journal of Electronic Materials}\ }\textbf
		{\bibinfo {volume} {38}},\ \bibinfo {pages} {1412} (\bibinfo {year}
		{2009})}\BibitemShut {NoStop}%
	\bibitem [{\citenamefont {{\AA}ngqvist}\ \emph {et~al.}(2016)\citenamefont
		{{\AA}ngqvist}, \citenamefont {Lindroth},\ and\ \citenamefont
		{Erhart}}]{erhart2016}%
	\BibitemOpen
	\bibfield  {author} {\bibinfo {author} {\bibfnamefont {M.}~\bibnamefont
			{{\AA}ngqvist}}, \bibinfo {author} {\bibfnamefont {D.~O.}\ \bibnamefont
			{Lindroth}}, \ and\ \bibinfo {author} {\bibfnamefont {P.}~\bibnamefont
			{Erhart}},\ }\href {\doibase 10.1021/acs.chemmater.6b02117} {\bibfield
		{journal} {\bibinfo  {journal} {Chemistry of Materials}\ }\textbf {\bibinfo
			{volume} {28}},\ \bibinfo {pages} {6877} (\bibinfo {year}
		{2016})}\BibitemShut {NoStop}%
	\bibitem [{\citenamefont {Perdew}\ \emph {et~al.}(2008)\citenamefont {Perdew},
		\citenamefont {Ruzsinszky}, \citenamefont {Csonka}, \citenamefont {Vydrov},
		\citenamefont {Scuseria}, \citenamefont {Constantin}, \citenamefont {Zhon},\
		and\ \citenamefont {Burke}}]{perdew2008:PBEsol}%
	\BibitemOpen
	\bibfield  {author} {\bibinfo {author} {\bibfnamefont {J.~P.}\ \bibnamefont
			{Perdew}}, \bibinfo {author} {\bibfnamefont {A.}~\bibnamefont {Ruzsinszky}},
		\bibinfo {author} {\bibfnamefont {G.~I.}\ \bibnamefont {Csonka}}, \bibinfo
		{author} {\bibfnamefont {O.~A.}\ \bibnamefont {Vydrov}}, \bibinfo {author}
		{\bibfnamefont {G.~E.}\ \bibnamefont {Scuseria}}, \bibinfo {author}
		{\bibfnamefont {L.~A.}\ \bibnamefont {Constantin}}, \bibinfo {author}
		{\bibfnamefont {X.}~\bibnamefont {Zhon}}, \ and\ \bibinfo {author}
		{\bibfnamefont {K.}~\bibnamefont {Burke}},\ }\href@noop {} {\bibfield
		{journal} {\bibinfo  {journal} {Phys. Rev. Lett.}\ }\textbf {\bibinfo
			{volume} {100}},\ \bibinfo {pages} {136406} (\bibinfo {year}
		{2008})}\BibitemShut {NoStop}%
	\bibitem [{\citenamefont {Gulans}\ \emph {et~al.}(2014)\citenamefont {Gulans},
		\citenamefont {Kontur}, \citenamefont {Meisenbichler}, \citenamefont {Nabok},
		\citenamefont {Pavone}, \citenamefont {Rigamonti}, \citenamefont
		{Sagmeister}, \citenamefont {Werner},\ and\ \citenamefont
		{Draxl}}]{gulans2014}%
	\BibitemOpen
	\bibfield  {author} {\bibinfo {author} {\bibfnamefont {A.}~\bibnamefont
			{Gulans}}, \bibinfo {author} {\bibfnamefont {S.}~\bibnamefont {Kontur}},
		\bibinfo {author} {\bibfnamefont {C.}~\bibnamefont {Meisenbichler}}, \bibinfo
		{author} {\bibfnamefont {D.}~\bibnamefont {Nabok}}, \bibinfo {author}
		{\bibfnamefont {P.}~\bibnamefont {Pavone}}, \bibinfo {author} {\bibfnamefont
			{S.}~\bibnamefont {Rigamonti}}, \bibinfo {author} {\bibfnamefont
			{S.}~\bibnamefont {Sagmeister}}, \bibinfo {author} {\bibfnamefont
			{U.}~\bibnamefont {Werner}}, \ and\ \bibinfo {author} {\bibfnamefont
			{C.}~\bibnamefont {Draxl}},\ }\href@noop {} {\bibfield  {journal} {\bibinfo
			{journal} {Journal of Physics: Condensed Matter}\ }\textbf {\bibinfo {volume}
			{26}},\ \bibinfo {pages} {363202} (\bibinfo {year} {2014})}\BibitemShut
	{NoStop}%
	\bibitem [{com()}]{comdetails}%
	\BibitemOpen
	\href@noop {} {}\bibinfo {note} {See Supplemental Material which includes the
		Supplementary Figs. 1-5 and the References
		\cite{gulans2014,perdew2008:PBEsol,troppenz2017,kim2020,rigamonti2020,cellURL,wanglandau2001,schnabel2011}.}\BibitemShut
	{Stop}%
	\bibitem [{rig()}]{rigamonti2020}%
	\BibitemOpen
	\href@noop {} {}\bibinfo {note} {S. Rigamonti {\textit{et al.}},
		{\texttt{CELL}}: a python package for cluster expansion with a focus on
		complex alloys, in preparation.}\BibitemShut {Stop}%
	\bibitem [{cel()}]{cellURL}%
	\BibitemOpen
	\href@noop {} {}\bibinfo {note} {{\texttt{CELL}} documentation:
		{\url{https://sol.physik.hu-berlin.de/cell}}}\BibitemShut {NoStop}%
	\bibitem [{\citenamefont {Metropolis}\ \emph {et~al.}(1953)\citenamefont
		{Metropolis}, \citenamefont {Rosenbluth}, \citenamefont {Rosenbluth},
		\citenamefont {Teller},\ and\ \citenamefont {Teller}}]{metropolis1953}%
	\BibitemOpen
	\bibfield  {author} {\bibinfo {author} {\bibfnamefont {N.}~\bibnamefont
			{Metropolis}}, \bibinfo {author} {\bibfnamefont {A.~W.}\ \bibnamefont
			{Rosenbluth}}, \bibinfo {author} {\bibfnamefont {M.~N.}\ \bibnamefont
			{Rosenbluth}}, \bibinfo {author} {\bibfnamefont {A.~H.}\ \bibnamefont
			{Teller}}, \ and\ \bibinfo {author} {\bibfnamefont {E.}~\bibnamefont
			{Teller}},\ }\href {\doibase 10.1063/1.1699114} {\bibfield  {journal}
		{\bibinfo  {journal} {The Journal of Chemical Physics}\ }\textbf {\bibinfo
			{volume} {21}},\ \bibinfo {pages} {1087} (\bibinfo {year}
		{1953})}\BibitemShut {NoStop}%
	\bibitem [{\citenamefont {Hastings}(1970)}]{hastings1970}%
	\BibitemOpen
	\bibfield  {author} {\bibinfo {author} {\bibfnamefont {W.~K.}\ \bibnamefont
			{Hastings}},\ }\href {\doibase 10.1093/biomet/57.1.97} {\bibfield  {journal}
		{\bibinfo  {journal} {Biometrika}\ }\textbf {\bibinfo {volume} {57}},\
		\bibinfo {pages} {97} (\bibinfo {year} {1970})}\BibitemShut {NoStop}%
	\bibitem [{\citenamefont {Bragg}\ and\ \citenamefont
		{Williams}(1934)}]{bragg1934}%
	\BibitemOpen
	\bibfield  {author} {\bibinfo {author} {\bibfnamefont {W.~L.}\ \bibnamefont
			{Bragg}}\ and\ \bibinfo {author} {\bibfnamefont {E.~K.}\ \bibnamefont
			{Williams}},\ }\href {https://doi.org/10.1098/rspa.1934.0132} {\bibfield
		{journal} {\bibinfo  {journal} {Proc. Roy. Soc. A}\ }\textbf {\bibinfo
			{volume} {145}} (\bibinfo {year} {1934})}\BibitemShut {NoStop}%
	\bibitem [{\citenamefont {Bethe}\ and\ \citenamefont
		{Wills}(1935)}]{bethe1935}%
	\BibitemOpen
	\bibfield  {author} {\bibinfo {author} {\bibfnamefont {H.~A.}\ \bibnamefont
			{Bethe}}\ and\ \bibinfo {author} {\bibfnamefont {H.}~\bibnamefont {Wills}},\
	}\href {https://doi.org/10.1098/rspa.1935.0122} {\bibfield  {journal}
		{\bibinfo  {journal} {Proc. Roy. Soc. A}\ }\textbf {\bibinfo {volume} {150}}
		(\bibinfo {year} {1935})}\BibitemShut {NoStop}%
	\bibitem [{\citenamefont {Landau}\ and\ \citenamefont
		{Binder}(2014)}]{landau2014}%
	\BibitemOpen
	\bibfield  {author} {\bibinfo {author} {\bibfnamefont {D.~P.}\ \bibnamefont
			{Landau}}\ and\ \bibinfo {author} {\bibfnamefont {K.}~\bibnamefont
			{Binder}},\ }\enquote {\bibinfo {title} {A guide to {M}onte {C}arlo
			simulations in statistical physics},}\ \ (\bibinfo  {publisher} {Cambridge
		University Press},\ \bibinfo {year} {2014})\ Chap.\ \bibinfo {chapter}
	{4.2.3},\ \bibinfo {edition} {4th}\ ed.\BibitemShut {Stop}%
	\bibitem [{\citenamefont {Challa}\ \emph {et~al.}(1986)\citenamefont {Challa},
		\citenamefont {Landau},\ and\ \citenamefont {Binder}}]{challa1986}%
	\BibitemOpen
	\bibfield  {author} {\bibinfo {author} {\bibfnamefont {M.~S.~S.}\
			\bibnamefont {Challa}}, \bibinfo {author} {\bibfnamefont {D.~P.}\
			\bibnamefont {Landau}}, \ and\ \bibinfo {author} {\bibfnamefont
			{K.}~\bibnamefont {Binder}},\ }\href {\doibase 10.1103/PhysRevB.34.1841}
	{\bibfield  {journal} {\bibinfo  {journal} {Phys. Rev. B}\ }\textbf {\bibinfo
			{volume} {34}},\ \bibinfo {pages} {1841} (\bibinfo {year}
		{1986})}\BibitemShut {NoStop}%
	\bibitem [{\citenamefont {Binder}(1987)}]{binder1987}%
	\BibitemOpen
	\bibfield  {author} {\bibinfo {author} {\bibfnamefont {K.}~\bibnamefont
			{Binder}},\ }\href {\doibase 10.1088/0034-4885/50/7/001} {\bibfield
		{journal} {\bibinfo  {journal} {Reports on Progress in Physics}\ }\textbf
		{\bibinfo {volume} {50}},\ \bibinfo {pages} {783} (\bibinfo {year}
		{1987})}\BibitemShut {NoStop}%
	\bibitem [{\citenamefont {Binder}(1997)}]{binder1997}%
	\BibitemOpen
	\bibfield  {author} {\bibinfo {author} {\bibfnamefont {K.}~\bibnamefont
			{Binder}},\ }\href {http://stacks.iop.org/0034-4885/60/i=5/a=001} {\bibfield
		{journal} {\bibinfo  {journal} {Reports on Progress in Physics}\ }\textbf
		{\bibinfo {volume} {60}},\ \bibinfo {pages} {487} (\bibinfo {year}
		{1997})}\BibitemShut {NoStop}%
	\bibitem [{\citenamefont {Newman}\ and\ \citenamefont
		{Barkema}()}]{newman1999}%
	\BibitemOpen
	\bibfield  {author} {\bibinfo {author} {\bibfnamefont {M.~E.~J.}\
			\bibnamefont {Newman}}\ and\ \bibinfo {author} {\bibfnamefont {G.~T.}\
			\bibnamefont {Barkema}},\ }\href@noop {} {\emph {\bibinfo {title} {Monte
				Carlo Methods in Statistical Physics}}}\ (\bibinfo  {publisher} {Clarendon
		Press})\BibitemShut {NoStop}%
	\bibitem [{\citenamefont {Schnabel}\ \emph {et~al.}(2011)\citenamefont
		{Schnabel}, \citenamefont {Seaton}, \citenamefont {Landau},\ and\
		\citenamefont {Bachmann}}]{schnabel2011}%
	\BibitemOpen
	\bibfield  {author} {\bibinfo {author} {\bibfnamefont {S.}~\bibnamefont
			{Schnabel}}, \bibinfo {author} {\bibfnamefont {D.~T.}\ \bibnamefont
			{Seaton}}, \bibinfo {author} {\bibfnamefont {D.~P.}\ \bibnamefont {Landau}},
		\ and\ \bibinfo {author} {\bibfnamefont {M.}~\bibnamefont {Bachmann}},\
	}\href {\doibase 10.1103/PhysRevE.84.011127} {\bibfield  {journal} {\bibinfo
			{journal} {Phys. Rev. E}\ }\textbf {\bibinfo {volume} {84}},\ \bibinfo
		{pages} {011127} (\bibinfo {year} {2011})}\BibitemShut {NoStop}%
	\bibitem [{\citenamefont {Kim}\ \emph {et~al.}(2020)\citenamefont {Kim},
		\citenamefont {Gulans},\ and\ \citenamefont {Draxl}}]{kim2020}%
	\BibitemOpen
	\bibfield  {author} {\bibinfo {author} {\bibfnamefont {J.}~\bibnamefont
			{Kim}}, \bibinfo {author} {\bibfnamefont {A.}~\bibnamefont {Gulans}}, \ and\
		\bibinfo {author} {\bibfnamefont {C.}~\bibnamefont {Draxl}},\ }\href
	{\doibase 10.1088/2516-1075/ababde} {\bibfield  {journal} {\bibinfo
			{journal} {Electronic Structure}\ }\textbf {\bibinfo {volume} {2}},\ \bibinfo
		{pages} {037001} (\bibinfo {year} {2020})}\BibitemShut {NoStop}%
\end{thebibliography}
%

\end{document}